\documentclass[useAMS,usenatbib]{mn2e}
\usepackage[dvips]{graphicx}
\usepackage{epsfig,amssymb,amsmath,color}
\newcommand{\beq}{\begin{equation}}
\newcommand{\eeq}{\end{equation}}
\newcommand{\beqn}{\begin{eqnarray}}
\newcommand{\eeqn}{\end{eqnarray}}

\def\bmath#1{\mbox{\boldmath$#1$}}

\long\def\symbolfootnote[#1]#2{\begingroup%
\def\thefootnote{\fnsymbol{footnote}}\footnote[#1]{#2}\endgroup}

\usepackage{pstricks,pst-text}

\title[Distributed Radio Interferometric Calibration]{Distributed Radio Interferometric Calibration}
\author[Yatawatta]{Sarod Yatawatta\\
ASTRON, Postbus 2, 7990 AA Dwingeloo, the Netherlands}

\begin{document}
\date{\today}
\pagerange{\pageref{firstpage}--\pageref{lastpage}} \pubyear{2015}
\maketitle
\label{firstpage}
\begin{abstract}
Increasing data volumes delivered by a new generation of radio interferometers require { computationally efficient} and robust calibration algorithms. In this paper, we propose distributed calibration as a way of improving both computational { cost} as well as robustness in calibration. We exploit the data parallelism across frequency that is inherent in radio astronomical observations that are recorded as multiple channels { at} different frequencies. Moreover, we also exploit the smoothness of the variation of calibration parameters across frequency. Data parallelism enables us to distribute the computing load across a network of compute agents. Smoothness in frequency enables us reformulate calibration as a consensus optimization problem. With this formulation, we enable flow of information between compute agents calibrating data at different frequencies, without actually passing the data, { and thereby improving robustness}. We present simulation results to show the feasibility as well as the advantages of distributed calibration as opposed to conventional calibration.
\end{abstract}
\section{Introduction}
Many of the science drivers in modern radio astronomy seek weak signals buried in noise and bright foregrounds. Existing radio interferometers are upgraded and new ones are { being} built to deliver large volumes of data to achieve this goal. A major step of data processing in such telescopes is the correction of systematic errors and the removal of contaminating foregrounds from the data, which is  called {\em calibration}. With wide fields of view, calibration has to be done along hundreds of directions in the sky, especially at low radio frequencies \citep{Bregman}. This entails solving for a large number of unknowns and a reliable solution can only be obtained if there are sufficient constraints. 

  There are several novel calibration algorithms \citep{Kaz2,Kaz3,ICASSP13,Tasse} that are presently being used that improve speed and robustness in calibration. Most of these algorithms \citep{Kaz2,Kaz3,ICASSP13} use an algebraic data model that directly solve for Jones matrices representing the cumulative effect of the systematic errors. On the other hand, solving for a physical model \citep{Bregman,Tasse} would reduce the number of unknowns, especially since most of the systematic errors are known to have a smooth variation across frequency. One drawback of the physical model based calibration is the need to access data across a wide frequency range, which is computationally not feasible { at a central location} given that there are thousands of frequency channels in the data. Moreover, a physical model requires an accurate description of the frequency dependence and this can only be done for specific { and well studied} errors. Therefore, in this paper we propose a distributed calibration scheme that preserve the simplicity and computational speed of algebraic model based calibration while enforcing the smoothness of the calibration parameters across frequency. This can be thought of as getting the best of both aforementioned calibration approaches. In order to do this, we reformulate calibration as a distributed optimization problem and use {\em consensus optimization} \citep{boyd2011}. 

Distributed learning and distributed optimization \citep{Tsitsiklis,BT} is a widely researched topic in various disciplines. Consensus optimization  \citep{boyd2011} is one algorithm for distributed learning. In the era of exascale computing and big data, the importance of such algorithms grow ever more (for some recent  results see for instance \cite{Chang2014,Wei2012,Mota2013}). Instead of one compute agent accessing data across all frequencies (which is computationally unfeasible), we consider a situation where a group of compute agents accessing data across smaller frequency intervals. This matches ideally with how radio astronomical data is organized (data for the full observing bandwidth is divided into channels and channels are grouped into subbands), { and stored}. Therefore, we consider a situation where each compute agent having access to only a few subbands (while the full bandwidth consists of a few hundred subbands). Each compute agent will calibrate the data available locally (using an algebraic model) and the calibration solutions are transferred to a centralized location (fusion center). At the fusion center, consensus on the smoothness of the parameters across frequency is enforced. { Afterwards,} this update is passed back to each compute agent. Therefore, indirectly, each compute agent receives information across the whole frequency range, thus improving the calibration. Moreover, since no attempt is made to directly model or estimate underlying physical parameters, the calibration algorithms are simpler and less susceptible to model errors. Furthermore, the amount of information that needs to be exchanged between the fusion center and the compute agents is much less compared to the amount of data being calibrated, making this scheme computationally feasible. We also note that similar approaches have been proposed and tested for radio astronomical image synthesis { \citep{Ferrari2014,PURIFY}} to reduce the number of Fourier space samples used in imaging as well as to improve the quality of reconstruction. { Such imaging approaches would certainly complement the calibration approach proposed in this paper.}

The rest of the paper is organized as follows: In section \ref{sec:model}, we introduce radio interferometric calibration and in section \ref{sec:distcal}, we reformulate it as a distributed consensus optimization problem. We give results based on simulations in section \ref{sec:simul} to show the feasibility and superiority of the proposed scheme and draw our conclusions in section \ref{sec:conclusions}.

{\em Notation}: Lower case bold letters refer to column vectors (e.g. ${\bmath y}$). Upper case bold letters refer to matrices (e.g. ${\bf {\sf C}}$). Unless otherwise stated, all parameters are complex numbers. The set of complex numbers is given as ${\mathbb C}$ while the set of real numbers as  ${\mathbb R}$. The matrix pseudo-inverse, transpose, and Hermitian transpose are referred to as $(.)^{\dagger}$, $(.)^{T}$, $(.)^{H}$, respectively. The matrix Kronecker product is given by $\otimes$. The identity matrix is given by ${\bf {\sf I}}$.  Estimated parameters are denoted by a hat, $\widehat{(.)}$. The Frobenius norm is given by $\|.\|$. A uniform distribution in $[0,1]$ is given as $\mathcal{U}(0,1)$.

\section{Radio Interferometric Calibration}\label{sec:model}
In this section we give a brief overview of the data model used in radio interferometric calibration \citep{HBS,TMS}. We consider the radio frequency sky that is part of the sky model to be composed of discrete sources, far away from the earth such that the approaching radiation from each one of them appears to be plane waves. However, in reality there is large scale diffuse structure as well. There are $N$ receiving elements with dual polarized feeds in the array and  at the $p$-th station, this plane wave causes an induced voltage, which is dependent on the beam attenuation as well as the radio frequency receiver chain attenuation. Consider the correlation of signals at the $p$-th receiver and the $q$-th receiver, with proper signal delay at frequency $f$ and time $t$ (with finite bandwidth and integration time). After correlation, the correlated signal of the $p$-th station and the $q$-th station (named as the {\em visibilities}), ${\bf {\sf V}}(p,q,t,f)$ ($\in {\mathbb C}^{2\times 2}$) is given by 
\beq \label{vispq}
{\bf {\sf V}}(p,q,t,f)= \sum_{k=1}^{K}  {\bf {\sf J}}(p,k,t,f) {\bf {\sf C}}(p,q,k,t,f) {\bf {\sf J}}(q,k,t,f)^{H} + {\bf {\sf N}}_{pq}.
\eeq
In (\ref{vispq}), ${\bf {\sf J}}(p,k,t,f)$ and ${\bf {\sf J}}(q,k,t,f)$ are the Jones matrices describing errors along the direction of source $k$, at stations $p$ and $q$, at time $t$ and frequency $f$, respectively. These matrices represent the effects of the propagation medium, the beam shape and the receiver. There are $K$ sources in the sky model and the noise matrix is given as ${\bf {\sf N}}_{pq}$ ($\in {\mathbb C}^{2\times 2}$). The contribution from the $k$-th source on baseline $pq$ is given by the coherency matrix ${\bf {\sf C}}(p,q,k,t,f)$ ($\in {\mathbb C}^{2\times 2}$). We consider the sources in the sky model to be unpolarized and for the $k$-th direction, with intensity $I(p,q,k,f)$ (invariant over time { but dependent on $p$,$q$ if the source is resolved}) we have 
\beq \label{coh}
{\bf {\sf C}}(p,q,k,t,f)=e^{\jmath \phi(p,q,k,t,f)}\left[ \begin{array}{cc}
I(p,q,k,f) & 0\\
0 & I(p,q,k,f)
\end{array} \right]
\eeq
where $\phi(p,q,k,t,f)$ is the Fourier phase component that depends on the direction in the sky as well as the separation of stations $p$ and $q$ and can be exactly calculated. Moreover, it is also possible to refine ${\bf {\sf C}}(p,q,k,t,f)$ to include finite integration time and bandwidth \citep{TMS} but in this paper we use the simpler form. 
 The noise matrix ${\bf {\sf N}}_{pq}$ is normally assumed to have elements with zero mean, complex Gaussian entries with equal variance in real and imaginary parts but the statistics will vary because of the unmodeled structure \citep{Kaz3}.

Calibration is the estimation of a set of parameters ${\bmath \theta}$ that describe the Jones matrices ${\bf {\sf J}}(p,k,t,f)$ for $p\in[1,N]$ and $k\in[1,K]$ for given $t$ and $f$. The solutions obtained are additionally used to correct the data  and also to calculate the residual by subtracting the predicted model from the (corrected) data. The maximum likelihood (ML) estimate of ${\bmath \theta}$ under zero mean, white Gaussian noise is obtained by minimizing the least squares cost function
\beqn \label{mltheta}
\lefteqn{g({\bmath \theta})=\sum_{t,f} \sum_{p,q} \|\ \  {\bf {\sf V}}(p,q,t,f)} && \\\nonumber
&& -\sum_{k=1}^{K}{\bf {\sf J}}(p,k,t,f) {\bf {\sf C}}(p,q,k,t,f) {\bf {\sf J}}(q,k,t,f)^{H}\ \ \|^2
\eeqn
and can be improved by using a weighted least squares estimator to account for errors in the sky model \citep{Kaz3}. At this point, we make several points clear and make certain assumptions:
\begin{itemize}
\item The solutions ${\bmath \theta}$ are assumed to be invariant over time, within the time interval $g({\bmath \theta})$ is minimized, therefore from now on, we drop the time dependence from ${\bf {\sf J}}(p,k,t,f)$ and use ${\bf {\sf J}}(p,k,f)$ instead. This can also be done for ${\bf {\sf C}}(p,q,k,t,f)$  to have ${\bf {\sf C}}(p,q,k,f)$ { and ${\bf {\sf V}}(p,q,t,f)$ to have ${\bf {\sf V}}(p,q,f)$, because the geometry} of baseline $pq$ is dependent on $t$ (also the summation over $t$ is implicitly assumed but not explicitly stated).
\item The solutions for different directions $k$ are assumed to have statistically independent noise, therefore, we use expectation maximization (EM) and space alternating expectation maximization (SAGE) \citep{Fess94,Kaz2} to simplify the cost function in (\ref{mltheta}) over summation in $k$.
\item We {\em do} assume variability of the solutions over $f$, indeed, this is the novelty of this paper. For some directions, we can assume a smooth variation  of the underlying parameters over $f$ as done in \citep{Tasse}. However, the drawback of the approach taken in \citep{Tasse} is the amount of data needed to get a reliable estimate of this variation over $f$.  Indeed for a telescope like LOFAR that observe over a wide bandwidth, producing hundreds of subbands and thousands of channels of data, even storing all subbands at one location is problematic, let alone reading that data into memory.
\end{itemize}

Therefore, in this paper, we reformulate calibration as a distributed optimization problem. We assume smooth variation of the parameters over $f$, but unlike in \citep{Tasse}, we do not directly estimate those underlying parameters. In other words, the smooth variation is imposed as an additional constraint, but the calibration problem is still kept unchanged by estimating ${\bf {\sf J}}(p,k,f)$ for each $p$, $k$ and $f$. Note that since the storage of data is { by default} distributed over $f$, i.e. different subbands (channels) are stored at different locations, the optimization can also be done in a distributed way. { This distribution of computations does not necessarily reduce the total computational cost, but it can reduce the computational cost required at any one location where data is stored, provided that the computations only access the data that is locally available. In the following section, we describe how this can be done.}

\section{Distributed calibration}\label{sec:distcal}
Consider the Jones matrices along the $k$-th direction, ${\bf {\sf J}}(p,k,f)$, for $N$ stations, let
\beq \label{Jdef}
{\bf {\sf J}}_{kf} \buildrel \triangle \over =\left[ \begin{array}{c}
 {\bf {\sf J}}(1,k,f)\\
 {\bf {\sf J}}(2,k,f)\\
 \vdots\\
 {\bf {\sf J}}(N,k,f)
\end{array} \right]
\eeq
where ${\bf {\sf J}}_{kf}$ ($\in {\mathbb C}^{2N\times 2}$) is the augmented Jones matrix. Also define a canonical selection matrix ${\bf {\sf A}}_p$ ($\in {\mathbb C}^{2\times 2N}$)
\beq \label{Ap}
{\bf {\sf A}}_p \buildrel\triangle\over=[{\bf 0},{\bf 0},\ldots,{\bf {\sf I}},\ldots,{\bf 0}].
\eeq
where all elements of ${\bf {\sf A}}_p$ are zero except the $p$-th block which is an identity matrix. Using ${\bf {\sf A}}_p$ and ${\bf {\sf J}}_{kf}$, we can recover ${\bf {\sf J}}(p,k,f)= {\bf {\sf A}}_p {\bf {\sf J}}_{kf}$.

{
The ML estimate for ${\bmath \theta}$ can ideally be obtained by minimizing (\ref{mltheta}), but this needs access to all data. In {\em normal} calibration, solutions are obtained separately for each $f$, using data at that frequency. For given $f$, consider partitioning the parameters as $\{{\bmath \theta}_{kf}: k=1\ldots K\}$. We apply the EM/SAGE algorithm \citep{Kaz2} to estimate each ${\bmath \theta}_{kf}$. The {\em expectation} step in SAGE finds the visibility contribution ${\bf {\sf V}}_{pqkf}$  from ${\bf {\sf V}}(p,q,f)$ (with ${\bf {\sf C}}_{pqkf}={\bf {\sf C}}(p,q,k,f)$) as
\beq \label{sum1}
{\bf {\sf V}}_{pqkf}={\bf {\sf V}}(p,q,f)-\sum_{l, l \ne k} {\bf {\sf A}}_p {\bf {\sf J}}_{lf} {\bf {\sf C}}_{pqlf} \left({\bf {\sf A}}_q {\bf {\sf J}}_{lf}\right)^H 
\eeq
and using this, in the {\em maximization} step, the current estimate for ${\bmath \theta}_{kf}$ is obtained by minimizing
\beq \label{sum2}
g_{kf}({\bmath \theta}_{kf})=\sum_{p,q} \|{\bf {\sf V}}_{pqkf}-{\bf {\sf A}}_p {\bf {\sf J}}_{kf} {\bf {\sf C}}_{pqkf} \left({\bf {\sf A}}_q {\bf {\sf J}}_{kf}\right)^H \|^2.
\eeq
}

Now, to simplify the description even further, we only consider calibration along the  $k$-th direction { or minimizing $g_{kf}({\bmath \theta}_{kf})$}, so we drop the subscript $k$. Let ${\bmath \theta}_{kf}={\bf {\sf J}}_{kf}={\bf {\sf J}}_{f}$ where ${\bf {\sf J}}_{f}$ is defined as in (\ref{Jdef}). Thereafter, we have the simplified form for (\ref{sum2})
\beq \label{sum3}
g_f({\bf {\sf J}}_{f})=\sum_{p,q} \| {\bf {\sf V}}_{pqf} - {\bf {\sf A}}_p {\bf {\sf J}}_{f} {\bf {\sf C}}_{pqf} \left({\bf {\sf A}}_q {\bf {\sf J}}_{f}\right)^H \|^2.
\eeq
So far, we have not imposed the smoothness over $f$ to ${\bf {\sf J}}_{f}$, in order to do that, we introduce hidden variables ${\bf {\sf Z}}_{l}$ ($\in {\mathbb C}^{2N\times 2}$), $l\in [1,F]$, and we enforce the relationship
\beq \label{hidden}
{\bf {\sf J}}_{f}=\sum_l b_l(f) {\bf {\sf Z}}_{l}
\eeq 
onto ${\bf {\sf J}}_{f}$. In (\ref{hidden}), the only frequency dependence on the right hand side is introduced by real scalar values $b_l(f)$ that can be thought of as polynomial terms (in $f$). The order of the polynomial is $F-1$ (where $F>1$) and this controls the smoothness. For instance, given reference frequency $f_0$, we can select $b_l(f)=\left(\frac{f-f_0}{f_0}\right)^{l-1}$, but this is one possible polynomial and we can use more sophisticated expressions if needed.

\begin{figure}
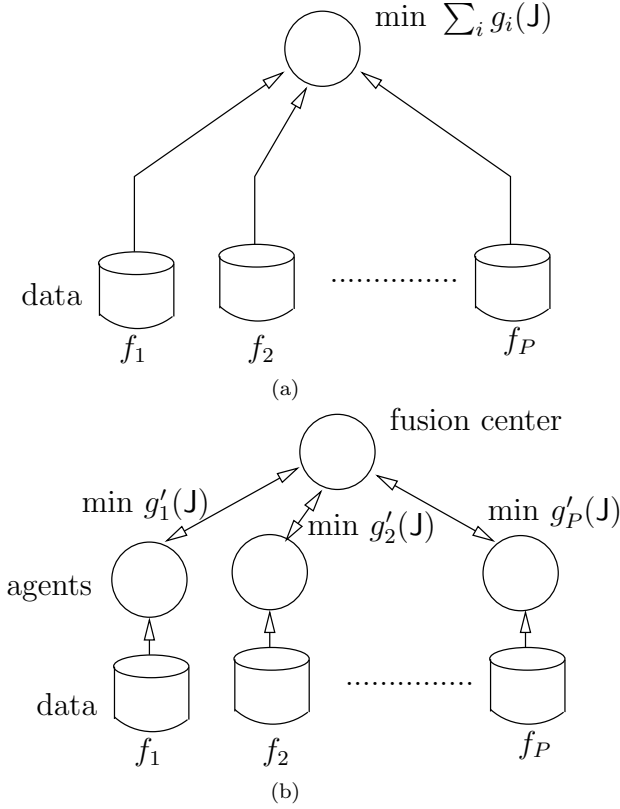

\begin{minipage}{0.98\linewidth}
\begin{minipage}{0.98\linewidth}
\begin{center}
\input{centralized.pstex_t}\\
(a)
\end{center}
\end{minipage}
\begin{minipage}{0.98\linewidth}
\begin{center}
\input{distributed.pstex_t}\\
(b)
\end{center}
\end{minipage}
\end{minipage}
\caption{Centralized calibration compared with distributed calibration. (a) Centralized calibration requires access to data observed at multiple frequencies. (b) Distributed calibration uses agents that operate on data taken at only a single frequency but via a fusion center, information is passed to other agents operating on data at different frequencies. The exact functions minimized in centralized calibration $g_i({\bf {\sf J}})$ and distributed calibration  $g_i^{\prime}({\bf {\sf J}})$ are slightly different. In normal calibration, each agent in (b) operate independently without communicating with the fusion center or any other agent.\label{block}}
\end{figure}

If ${\bmath b}_f$ ($\in {\mathbb R}^{F\times 1}$) is a vector representing all polynomial terms
\beq \label{poly}
{\bmath b}_f=[b_1(f)\  b_2(f)\  \ldots\  b_F(f)]^{T}
\eeq 
we can rewrite (\ref{hidden}) as
\beq \label{hblock}
{\bf {\sf J}}_{f}=\left( {\bmath b}_f^T \otimes {\bf {\sf I}}_{2N} \right) {\bf {\sf Z}} = {\bf {\sf B}}_{f}  {\bf {\sf Z}}
\eeq
where ${\bf {\sf I}}_{2N}$ is the $2N \times 2N$ identity matrix, ${\bf {\sf B}}_{f}=\left( {\bmath b}_f^T \otimes {\bf {\sf I}}_{2N} \right)$ ($\in {\mathbb R}^{2N\times 2FN}$)  and ${\bf {\sf Z}}$ ($\in {\mathbb C}^{2FN\times 2}$) is the augmented matrix of hidden variables
\beq
{\bf {\sf Z}}=[{\bf {\sf Z}}_{1}^T\ {\bf {\sf Z}}_{2}^T \ldots {\bf {\sf Z}}_{F}^T]^T.
\eeq
For each direction $k$, by imposing smoothness, we can find a set of hidden variables ${\bf {\sf Z}}$ for any given value for $F$. At this point, we distinguish between direct estimation of ${\bf {\sf Z}}$ and the method proposed in this paper:
\begin{itemize}
\item Centralized calibration is estimating ${\bf {\sf Z}}$ directly from the data. However, this requires access to all frequencies (or at least a set of frequencies more than $F$) as shown in Fig. \ref{block} (a). More rigorously, centralized calibration is estimating ${\bf {\sf Z}}$ such that $\sum_f g_f({\bf {\sf J}}_{f})$ is minimized. As we explained before, this is computationally not feasible because of the large data volumes needed.
\item Instead of centralized calibration, we formulate distributed calibration as follows. Let there be $P$ computational agents (or nodes) in a network. We assume the $i$-th agent will only have access to the data at frequency $f_i$ as in Fig. \ref{block} (b). However, we enforce {\em consensus} among all agents, in other words, we enforce an additional constraint ${\bf {\sf J}}_{f_i}= {\bf {\sf B}}_{f_i} {\bf {\sf Z}}$ that all agents have to satisfy. { Note that the total number of frequencies that the data is taken will almost surely be higher than $P$. In that case, we consider calibration of a subset of $P$ frequencies selected from the total available frequencies. This selection has to be repeated sequentially until all the frequencies are calibrated.}
\end{itemize}

With this network setup, we formulate distributed calibration as
\beqn
\{{\bf {\sf J}}_{f_1},{\bf {\sf J}}_{f_2},\ldots,{\bf {\sf Z}}\}=\underset{{\bf {\sf J}}_{f_i},\ldots,{\bf {\sf Z}}}{\rm arg\ min} \sum_i g_{f_i}({\bf {\sf J}}_{f_i})\\\nonumber
{\rm subject\ to}\ \  {\bf {\sf J}}_{f_i}={\bf {\sf B}}_{f_i} {\bf {\sf Z}},\ \ i\in[1,P]
\eeqn
which is actually a consensus optimization problem \citep{boyd2011}. To solve this, we use the augmented Lagrangian method with the Lagrangian
\beqn \label{AugL}
\lefteqn{
L({\bf {\sf J}}_{f_1},{\bf {\sf J}}_{f_2},\ldots,{\bf {\sf Z}},{\bf {\sf Y}}_{f_1},{\bf {\sf Y}}_{f_2},\ldots) }\\\nonumber
&& =\sum_i g_{f_i}({\bf {\sf J}}_{f_i}) + \| {\bf {\sf Y}}_{f_i}^H ({\bf {\sf J}}_{f_i}- {\bf {\sf B}}_{f_i} {\bf {\sf Z}})\| + \frac{\rho}{2} \| {\bf {\sf J}}_{f_i}- {\bf {\sf B}}_{f_i} {\bf {\sf Z}} \|^2 \\\nonumber
&& =\sum_i L_i\left({\bf {\sf J}}_{f_i},{\bf {\sf Z}},{\bf {\sf Y}}_{f_i}\right)
\eeqn
where ${\bf {\sf Y}}_{f_i}$ are the Lagrange multipliers and $\rho$ is the regularization factor. In order to solve (\ref{AugL}), we use the consensus alternating direction method of multipliers (C-ADMM) \citep{boyd2011}. If superscript $n$ denote values at the $n$-th C-ADMM iteration, the values for the $(n+1)$-th iteration are updated as
\beqn \label{M1}
({\bf {\sf J}}_{f_i})^{n+1}=\underset{{\bf {\sf J}}}{\rm min}\  L_i\left({\bf {\sf J}},({\bf {\sf Z}})^n,({\bf {\sf Y}}_{f_i})^n\right)\\ \label{M2}
({\bf {\sf Z}})^{n+1}=\underset{{\bf {\sf Z}}}{\rm min} \sum_i  L_i\left(({\bf {\sf J}}_{f_i})^{n+1},{\bf {\sf Z}},({\bf {\sf Y}}_{f_i})^n\right)\\ \label{M3}
({\bf {\sf Y}}_{f_i})^{n+1}=({\bf {\sf Y}}_{f_i})^{n} + \rho (({\bf {\sf J}}_{f_i})^{n+1}- {\bf {\sf B}}_{f_i} ({\bf {\sf Z}})^{n+1}).
\eeqn
The minimization (\ref{M1}) has no closed form solution and needs to be done iteratively, for instance by using the Broyden-Fletcher-Goldfarb-Shanno (BFGS) algorithm \citep{NW} or by using trust-region algorithms. In this paper, we use the Riemannian trust-region algorithm (RTR) described in \cite{RTR} for this minimization, and we need to calculate the gradient and the Hessian. The gradient and the Hessian with respect to ${\bf {\sf J}}_{f_i}$ of (\ref{M1}) are given as (see \cite{ICASSP13} and appendix \ref{app:grad} for proof)
\beq \label{gradL}
{\rm grad}_i(L_i,{\bf {\sf J}}) = {\rm grad}_i\left(g_{f_i}({\bf {\sf J}}),{\bf {\sf J}}\right) + \frac{1}{2} {\bf {\sf Y}}_{f_i} + \frac{\rho}{2} \left({\bf {\sf J}}- {\bf {\sf B}}_{f_i} {\bf {\sf Z}}\right)
\eeq
and 
\beq \label{HessL}
{\rm Hess}_i(L_i,{\bf {\sf J}},{\bmath \eta}) = {\rm Hess}_i\left(g_{f_i}({\bf {\sf J}}),{\bf {\sf J}},{\bmath \eta}\right) + {\bf {\sf 0}} + \frac{\rho}{2} {\bmath \eta}
\eeq
where we use (\ref{sum3}) to get
\beqn
\lefteqn{{\rm grad}_i\left(g_{f_i}({\bf {\sf J}}),{\bf {\sf J}}\right)}\\\nonumber
&=&-\sum_{p,q} \left({\bf {\sf A}}_p^T({\bf {\sf V}}_{pqf_i}-{\bf {\sf A}}_p {\bf {\sf J}} {\bf {\sf C}}_{pqf_i} {\bf {\sf J}}^H {\bf {\sf A}}_q^T){\bf {\sf A}}_q{\bf {\sf J}}{\bf {\sf C}}_{pqf_i}^H \right.\\\nonumber
&& + \left. {\bf {\sf A}}_q^T({\bf {\sf V}}_{pqf_i}-{\bf {\sf A}}_p{\bf {\sf J}}{\bf {\sf C}}_{pqf_i}{\bf {\sf J}}^H{\bf {\sf A}}_q^T)^H {\bf {\sf A}}_p {\bf {\sf J}}{\bf {\sf C}}_{pqf_i}\right)
\eeqn
and
\beqn
\lefteqn{{\rm Hess}_i\left(g_{f_i}({\bf {\sf J}}),{\bf {\sf J}},{\bmath \eta}\right)}\\\nonumber
&=&\sum_{p,q}\left( {\bf {\sf A}}_p^T \left( ({\bf {\sf V}}_{pqf_i}-{\bf {\sf A}}_p{\bf {\sf J}}{\bf {\sf C}}_{pqf_i}{\bf {\sf J}}^H{\bf {\sf A}}_q^T) {\bf {\sf A}}_q {\bmath \eta}\right.\right.\\\nonumber
&& \left.\left.- {\bf {\sf A}}_p({\bf {\sf J}}{\bf {\sf C}}_{pqf_i} {\bmath \eta}^H + {\bmath \eta}{\bf {\sf C}}_{pqf_i}{\bf {\sf J}}^H) {\bf {\sf A}}_q^T{\bf {\sf A}}_q{\bf {\sf J}}\right) {\bf {\sf C}}_{pqf_i}^H\right. \\\nonumber
&&\left. + {\bf {\sf A}}_q^T \left( ({\bf {\sf V}}_{pqf_i}-{\bf {\sf A}}_p{\bf {\sf J}}{\bf {\sf C}}_{pqf_i}{\bf {\sf J}}^H{\bf {\sf A}}_q^T)^H {\bf {\sf A}}_p {\bmath \eta}\right.\right.\\\nonumber
&& \left.\left.- {\bf {\sf A}}_q({\bf {\sf J}}{\bf {\sf C}}_{pqf_i} {\bmath \eta}^H + {\bmath \eta}{\bf {\sf C}}_{pqf_i}{\bf {\sf J}}^H)^H {\bf {\sf A}}_p^T{\bf {\sf A}}_p{\bf {\sf J}}\right) {\bf {\sf C}}_{pqf_i}\right). \\\nonumber
\eeqn

Minimization of (\ref{M2}) can be done in closed form. We take the derivative to get
\beq \label{gradZ}
{\rm grad}(L,{\bf {\sf Z}})=\sum_i {\bf {\sf B}}_{f_i}^T\left(-{\bf {\sf Y}}_{f_i} + \rho (-{\bf {\sf J}}_{f_i}+{\bf {\sf B}}_{f_i}{\bf {\sf Z}})\right)
\eeq
and equating this to zero gives us
\beq
{\bf {\sf Z}}=\left( \sum_i \rho {\bf {\sf B}}_{f_i}^T {\bf {\sf B}}_{f_i} \right)^{\dagger} \left(\sum_i {\bf {\sf B}}_{f_i}^T ({\bf {\sf Y}}_{f_i}+\rho {\bf {\sf J}}_{f_i})\right)
\eeq
which can be further simplified by using ${\bf {\sf B}}_{f_i}=\left( {\bmath b}_{f_i}^T \otimes {\bf {\sf I}}_{2N} \right)$ to get
\beq \label{Zest}
{\bf {\sf Z}}=\frac{1}{\rho}\left( \left(\sum_i {\bmath b}_{f_i} {\bmath b}_{f_i}^T\right)^{\dagger} \otimes {\bf {\sf I}}_{2N} \right) \left( \sum_i {\bmath b}_{f_i} \otimes ({\bf {\sf Y}}_{f_i}+\rho {\bf {\sf J}}_{f_i})\right).
\eeq
{ Each column of ${\bf {\sf Z}}$ in (\ref{Zest}) can be written as ${\bmath z} = \left( {\bf {\sf P}} \otimes {\bf {\sf I}}_{2N} \right) {\bmath r}$, where ${\bf {\sf P}}=\frac{1}{\rho}  \left(\sum_i {\bmath b}_{f_i} {\bmath b}_{f_i}^T\right)^{\dagger}$ ($\in {\mathbb R}^{F\times F}$) and ${\bmath r} \in {\mathbb C}^{2FN\times 1}$ is the corresponding column of the right hand sum of (\ref{Zest}). We can reshape ${\bmath r}$ to get $\widetilde{\bf {\sf R}} \in {\mathbb C}^{2N\times F}$. Then we can rewrite ${\bmath z} = {\mathrm{vec}}\left({\bf {\sf I}}_{2N} \widetilde{\bf {\sf R}} {\bf {\sf P}}^T\right) = \mathrm{vec}\left(\widetilde{\bf {\sf R}} {\bf {\sf P}}^T\right)$ which is far simpler to obtain than  directly solving (\ref{Zest}).}
Moreover, we see that from (\ref{Zest}) in order to have full rank, the summation $\sum_i {\bmath b}_{f_i} {\bmath b}_{f_i}^T$ should at least have $F$ terms (because the size of ${\bmath b}_{f_i} {\bmath b}_{f_i}^T$ is $F\times F$). In other words,  we need to have data for at least $F$ different frequencies, or $P\ge F$.

To recapitulate, we consider $P$ compute agents operating simultaneously. Each agent $i$ only has access to the data at frequency $f_i$. There is also a data fusion center with which each agent does communication. Consider calibration along a single direction first. Each agent $i$ needs to estimate ${\bf {\sf J}}_{f_i}$  ($\in {\mathbb C}^{2N\times 2}$) and will keep auxiliary variable ${\bf {\sf Y}}_{f_i}$  ($\in {\mathbb C}^{2N\times 2}$) locally. The fusion center will keep the global variable ${\bf {\sf Z}}$ ($\in {\mathbb C}^{2FN\times 2}$) and will pass the product ${\bf {\sf B}}_{f_i}{\bf {\sf Z}}$ ($\in {\mathbb C}^{2N\times 2}$) onto the $i$-th agent. With this additional variables, the C-ADMM algorithm for a single direction ($K=1$) can be described as:
\begin{enumerate}
\item [(S1)] Each agent $i$ finds estimate for ${\bf {\sf J}}_{f_i}$ by solving (\ref{M1}). Thereafter, it sends back the result ${\bf {\sf Y}}_{f_i} + \rho {\bf {\sf J}}_{f_i}$ to the fusion center.
\item [(S2)] At the fusion center, after collecting the values ${\bf {\sf Y}}_{f_i} + \rho {\bf {\sf J}}_{f_i}$ from all agents, (\ref{M2}) is minimized by solving (\ref{Zest}). Once the updated ${\bf {\sf Z}}$ is obtained, ${\bf {\sf B}}_{f_i}{\bf {\sf Z}}$ is sent back to the $i$-th agent.
\item [(S3)] At agent $i$, ${\bf {\sf Y}}_{f_i}$ is updated by using  (\ref{M3}) with the new value of ${\bf {\sf B}}_{f_i}{\bf {\sf Z}}$ received from the fusion center. If stopping criteria (such as the maximum C-ADMM iterations) are not met, we go back to (S1) above.
\end{enumerate}
Note that steps (S1) and (S3) above are done simultaneously at each agent. The centralized step (S2) is only an averaging step which is far less expensive compared with the minimization in (S1).

The above description is only for calibration along a single direction. In order to apply the same method for calibration along $K$ directions, we only need slight modifications to the steps described above. We use subscript $k$ to indicate the $k$-th direction. Each agent $i$ needs to estimate $K$ values ${\bf {\sf J}}_{kf_i}$. Moreover, each agent has $K$ auxiliary variables ${\bf {\sf Y}}_{kf_i}$.  The fusion center keeps the global variables ${\bf {\sf Z}}$ ($\in {\mathbb C}^{2KFN\times 2}$) which has $K$ blocks (let us denote the $k$-th block of ${\bf {\sf Z}}$ as $({\bf {\sf Z}})_k$), one for each direction. Therefore, we have the C-ADMM algorithm for $K$ directions as:
\begin{enumerate}
\item [(D1)] Each agent $i$ finds estimate for $K$ values ${\bf {\sf J}}_{kf_i}$. This is done by decomposing the $K$ direction problem onto $K$ problems of the type (\ref{M1}). In order to do this, we use SAGE algorithm \citep{Kaz2}. Note that in SAGE algorithm, we need to calculate the conditional mean of the data for each direction (expectation step) and we calculate this ignoring the auxiliary variables and the regularizing term. However, in the maximization step of the algorithm, we solve  (\ref{M1}) with full regularization. Thereafter, it sends back the results ${\bf {\sf Y}}_{kf_i} + \rho {\bf {\sf J}}_{kf_i}$ to the fusion center ($K$ values).
\item [(D2)] At the fusion center, The block matrix ${\bf {\sf Z}}$ is updated by solving (\ref{Zest}) for $K$ blocks separately. Thereafter, with the updated ${\bf {\sf Z}}$, ${\bf {\sf B}}_{f_i}({\bf {\sf Z}})_k$ is sent back to the $i$-th agent ($K$ values).
\item [(D3)] At agent $i$, ${\bf {\sf Y}}_{kf_i}$ for $K$ values are updated using (\ref{M3}) and if stopping criteria are not met, we go back to (D1) above.
\end{enumerate}

The initialization for the C-ADMM algorithm is done as follows. First, the initial values for ${\bf {\sf J}}_{kf_i}$ can be taken as a block matrix of $2\times 2$ identity matrices (or for a warm start, we can take the solutions from the previous time slot). The initial values for both ${\bf {\sf Z}}$ and ${\bf {\sf Y}}_{kf_i}$ are taken as ${\bf {\sf 0}}$. Because of this, the solutions obtained for ${\bf {\sf J}}_{kf_i}$ at the first C-ADMM iteration for steps (S1) and (D1)  will have an unknown unitary ambiguity \citep{interpolation}. Therefore, only at the first iteration, the averaging step in (S2) and (D2) should be done after projecting each ${\bf {\sf J}}_{kf_i}$ to the mean value calculated using the quotient manifold structure described in \citep{interpolation}. For the remaining iterations, because ${\bf {\sf Z}}$ and ${\bf {\sf Y}}_{kf_i}$ are not ${\bf {\sf 0}}$, the unitary ambiguity will be common (for each direction) and normal Euclidean averaging can be done in steps (S2) and (D2).

The selection of the regularization parameter $\rho$ ($>0$) is specific to each problem and more detail can be found in \cite{boyd2011}. We note here that it is possible to select different values of $\rho$ for different directions when $K$ directions are calibrated. For instance, for source clusters that are far away from the phase center, it might be true that there will not by any smooth variation of the errors with frequency along that direction. Therefore, for that specific direction, we can make $\rho$ very small (so no smoothness is enforced). On the other hand, for source clusters at the phase center (also at the center of the beam), we can safely assume that the errors vary smoothly with frequency and use a high value for $\rho$ (typically $\in[1,10]$). In section \ref{sec:simul}, we provide simulations where we have varied the value of $\rho$ and see how the performance change.

The convergence of distributed calibration is discussed in appendix \ref{app:conv} in detail. This boils down to { having} a sky model with finite, non-zero flux and data with finite values, { but the true sky can have zero flux, and then the solutions will be zero}. Convexity of the cost functions are also desired for C-ADMM to converge \citep{boyd2011} and for an interferometric data model, this generally is assumed to hold.

{ 
The amount of information that needs to be exchanged between the $i$-th agent and the fusion center is $K\times 2N\times 2$ (complex variables). In contrast, the amount of observed data used in calibration at the $i$-th agent is of the order $N(N-1)/2\times 2\times 2 \times T$ for $T$ time samples with $N(N-1)/2$ baselines. Therefore, when working with $P$ frequencies, the total amount of data that needs to be accessed is $T(N-1)/2 \times 4 NP$ and for $K$ directions, the total amount of information that needs to be exchanged in distributed calibration is $K \times 4 NP$. Hence, when $K \ll T (N-1)/2$, the amount of information passed is much less in distributed calibration, regardless of the value of $P$. 

The total number of computations in distributed calibration compared to normal calibration is not significantly different, and in fact it could even be higher. However, we gain a significant reduction in operational and energy cost by being able to distribute the total computations across a network of compute agents. In addition, there are several possibilities to reduce the computational cost even further and these will be explored in future work. First, it is possible to eliminate the need for having a fusion center \citep{Erseghe12,Wei2014} and design  an algorithm where agents only pass data between their neighbours. Secondly, when there are data with more frequencies than the number of compute agents, a multiplexing scheme where each agent alternates the data used in calibration, and yet calibrates the full dataset can be investigated.
}

\section{Simulations}\label{sec:simul}
In this section, we present several simulations to illustrate the performance of distributed calibration. We consider a radio telescope similar to LOFAR, observing in the frequency range 115 MHz to 185 MHz with $N=47$ stations pointed at the north celestial pole \citep{NCP2013}. We consider data taken at $P=32$ different channels (with bandwidth 0.2 MHz each) uniformly spaced in frequency within the observing frequency range. Therefore { the frequency range covered by the data is 70 MHz wide but the actual bandwidth is 6.4 MHz}. In a typical situation, in order to increase the number of constraints, calibration is performed for about every few minutes of data, using more than 1 time sample (for instance with 10 s integration, we have 30 samples for 5 minutes of data). { In our simulations we only use 20 time samples in all calibration tests, equivalent to a total integration time of 200 s}. Note that we call calibration of individual channels separately (without using the information across frequency) as {\em normal} calibration throughout this section.

We simulate (\ref{vispq}) and the Jones matrices ${\bf {\sf J}}(p,k,t,f)$ are simulated as follows. The variation with $t$ is simulated as $\sin(\alpha_1 t^\prime + 2\pi \beta_1) +\jmath\sin(\alpha_2 t^\prime + 2\pi \beta_2)$ where $\alpha_1,\alpha_2,\beta_1,\beta_2$ are drawn from a uniform distribution $\mathcal{U}(0,1)$ for each station $p$ and direction $k$, and $t^\prime$ is time sample number. The variation across frequency is simulated by using a polynomial $\sum_{l=1}^{G} (\gamma_l+\jmath \delta_l) \left(\frac{f-f_0}{f_0}\right)^{(l-1)}$ where $\gamma_l,\delta_l$ are also drawn from a uniform distribution $\mathcal{U}(0,1)$. The reference frequency $f_0$ is taken to be 150 MHz and $G=4$ for all simulations. The product of the time variation and frequency variation gives the complete description of the elements in ${\bf {\sf J}}(p,k,t,f)$.

The sky model consists of point sources, randomly distributed over a field of view of 7$\times$7 square degrees. Their intensities at frequency $f_0$ is simulated using a power law and their spectral indices are drawn from a uniform distribution $\mathcal{U}(-1,1)$. The flux of the weakest source calibrated is set to be 1 Jy. The number of sources $K$ simulated (and calibrated) is varied for different simulations as described below. In addition, we also simulate a set of 300 weak background sources, with peak flux below 0.1 Jy and have a flat spectrum. We do not corrupt these sources with errors because we want to examine the effect of calibration of the bright foreground sources on them.

Finally, we add noise ${\bf {\sf N}}_{pq}$ to the simulated visibilities in (\ref{vispq}). The elements of ${\bf {\sf N}}_{pq}$ are drawn from a Gaussian distribution with zero mean and equal variance in real and imaginary parts. The noise variance is adjusted such that the total noise power is 10\% of total signal power for the full observation, which is 6 hours.

\subsection{Simulation I}
We consider calibration along one direction $K=1$. For normal calibration, we use 30 iterations of the RTR algorithm { (beyond which we do not see any improvement)}. For distributed calibration, we use 50 C-ADMM iterations. Each C-ADMM iteration performs steps (S1,S2,S3) described in section \ref{sec:distcal} and step (S1) uses 10 iterations of the RTR algorithm. Let the simulated Jones matrices (\ref{Jdef}) at frequency $f$ be given by ${\bf {\sf J}}_f$ and its estimated value be given by $\widehat{\bf {\sf J}}_f$. Then the error between ${\bf {\sf J}}_f$ and $\widehat{\bf {\sf J}}_f$ (per parameter) is found as $\frac{1}{\sqrt{4N}}\|{\bf {\sf J}}_f-\widehat{\bf {\sf J}}_f {\bf {\sf U}}\|$ where ${\bf {\sf U}}$ ($\in {\mathbb C}^{2\times 2}$) is a unitary matrix denoting the unitary ambiguity between the true parameters and estimated parameters. It is found by solving a matrix Procrustes problem \citep{interpolation}. We average the error calculated this way over the $P$ frequencies and all time samples to get the final error.

{
Moreover, we use two measures of error to study the convergence of distributed calibration. We define the 'primal' residual as $\frac{1}{\sqrt{4N}}\|({\bf {\sf J}}_{f_i})^n-{\bf {\sf B}}_{f_i} ({\bf {\sf Z}})^n\|$, averaged over all $f_i$. The 'dual' residual is defined as $\frac{1}{\sqrt{4 F N}}\| ({\bf {\sf Z}})^{n+1}-({\bf {\sf Z}})^n \|$, where the superscripts $n+1$ and $n$ denote the C-ADMM iteration number. The primal residual depicts the error between the local solution and the predicted consensus value. On the other hand, the dual residual depicts the convergence of the global variable ${\bf {\sf Z}}$.

}
\begin{figure}
\begin{minipage}{0.98\linewidth}
\centering
 \centerline{\epsfig{figure=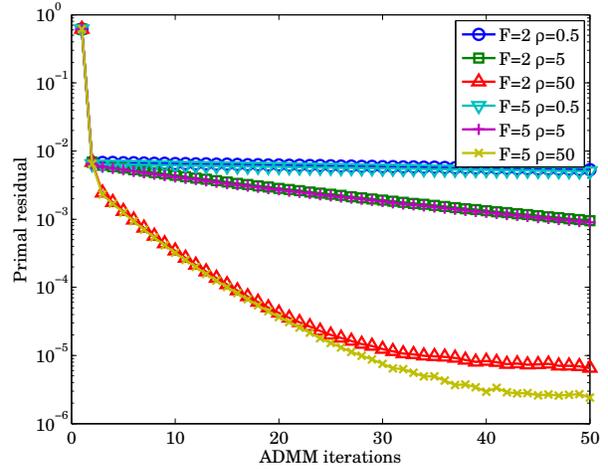,width=8.0cm}}
\caption{Variation of the primal residual with C-ADMM iteration number, for two values of smoothing polynomial terms $F=2$ and $F=5$ and three values of regularization factor $\rho=0.5$, $\rho=5$ and $\rho=50$.\label{primal_res}}
\end{minipage}
\end{figure}

\begin{figure}
\begin{minipage}{0.98\linewidth}
\centering
 \centerline{\epsfig{figure=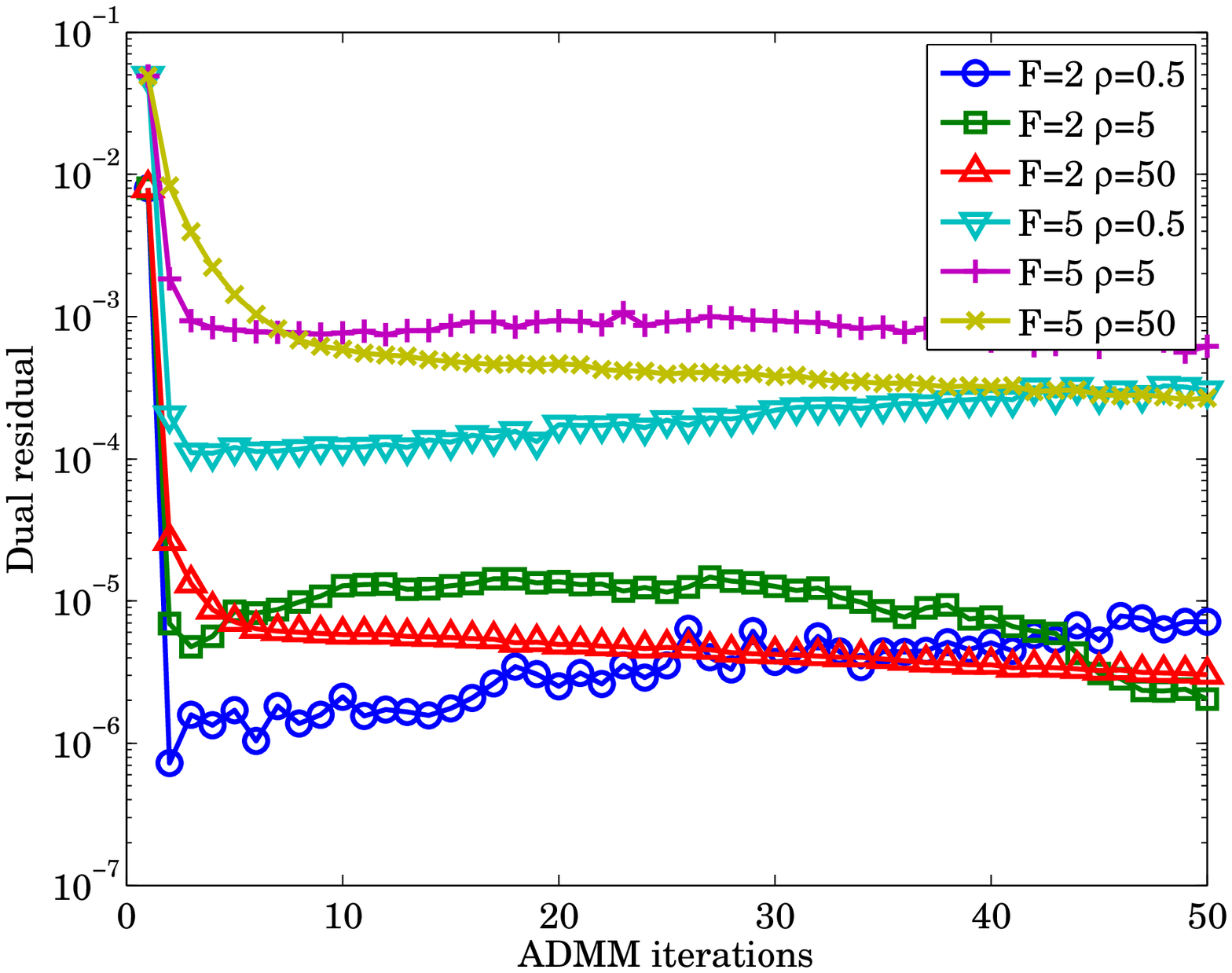,width=8.0cm}}
\caption{Variation of the dual residual with C-ADMM iteration number, for two values of smoothing polynomial terms $F=2$ and $F=5$ and three values of regularization factor $\rho=0.5$, $\rho=5$ and $\rho=50$.\label{dual_res}}
\end{minipage}
\end{figure}

\begin{figure}
\begin{minipage}{0.98\linewidth}
\centering
 \centerline{\epsfig{figure=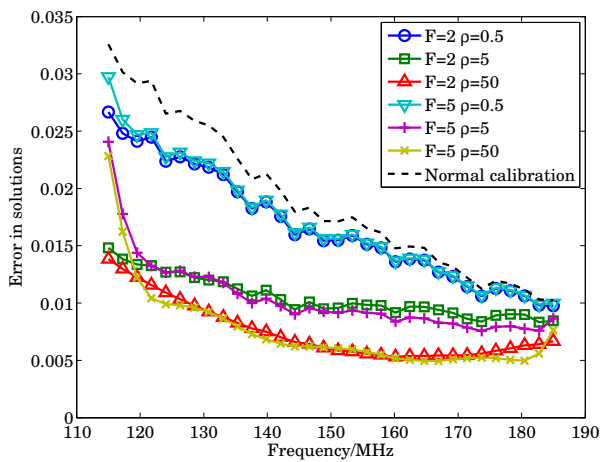,width=8.0cm}}
\caption{Variation of average error standard deviation of the estimated solutions with frequency after 50 C-ADMM iterations. Normal calibration has higher error and distributed calibration with $\rho=50$ gives the lowest error, both for $F=2$ and $F=5$. The edge frequencies have higher error for $F=5$ due to our choice of the interpolating polynomial.\label{errNF}}
\end{minipage}
\end{figure}

{
In Fig. \ref{primal_res}, we have shown the variation of the primal residual and in Fig. \ref{dual_res}, the variation of the dual residual, both with the C-ADMM iteration number. The regularization parameter is set at $\rho=0.5$, $\rho=5$ and $\rho=50$. The number of terms in the smoothing polynomial (\ref{poly}) is set at $F=2$ and $F=5$, with $F=2$ underestimating the simulated polynomial order while $F=5$ overestimating it. It is clear that as the value of $\rho$ increases, the primal residual converges faster, and to a lower value. Also, the dual residual is lower for a low order polynomial, or a lower value of $F$.

In Fig. \ref{errNF}, we show the average error per parameter for the chosen values of $F$ and $\rho$, after 50 C-ADMM iterations. In all cases, distributed calibration gives a lower error than normal calibration. Even though the true parameters are simulated using a polynomial with $G=4$, we get the lowest error for both $F=2$ and $F=5$, at $\rho=50$. The lower bound of this error is determined by the noise and the weak sources not included in calibration. For this example, this bound is too high to see a difference in performance between $F=2$ and $F=5$. Moreover, we also have errors due to polynomial interpolation, which is clearly seen for $F=5$ at the edge frequencies.
}

\subsection{Simulation II}
\begin{figure*}
\begin{minipage}{0.98\linewidth}
\begin{center}
\begin{minipage}{0.48\linewidth}
\centering
 \centerline{\epsfig{figure=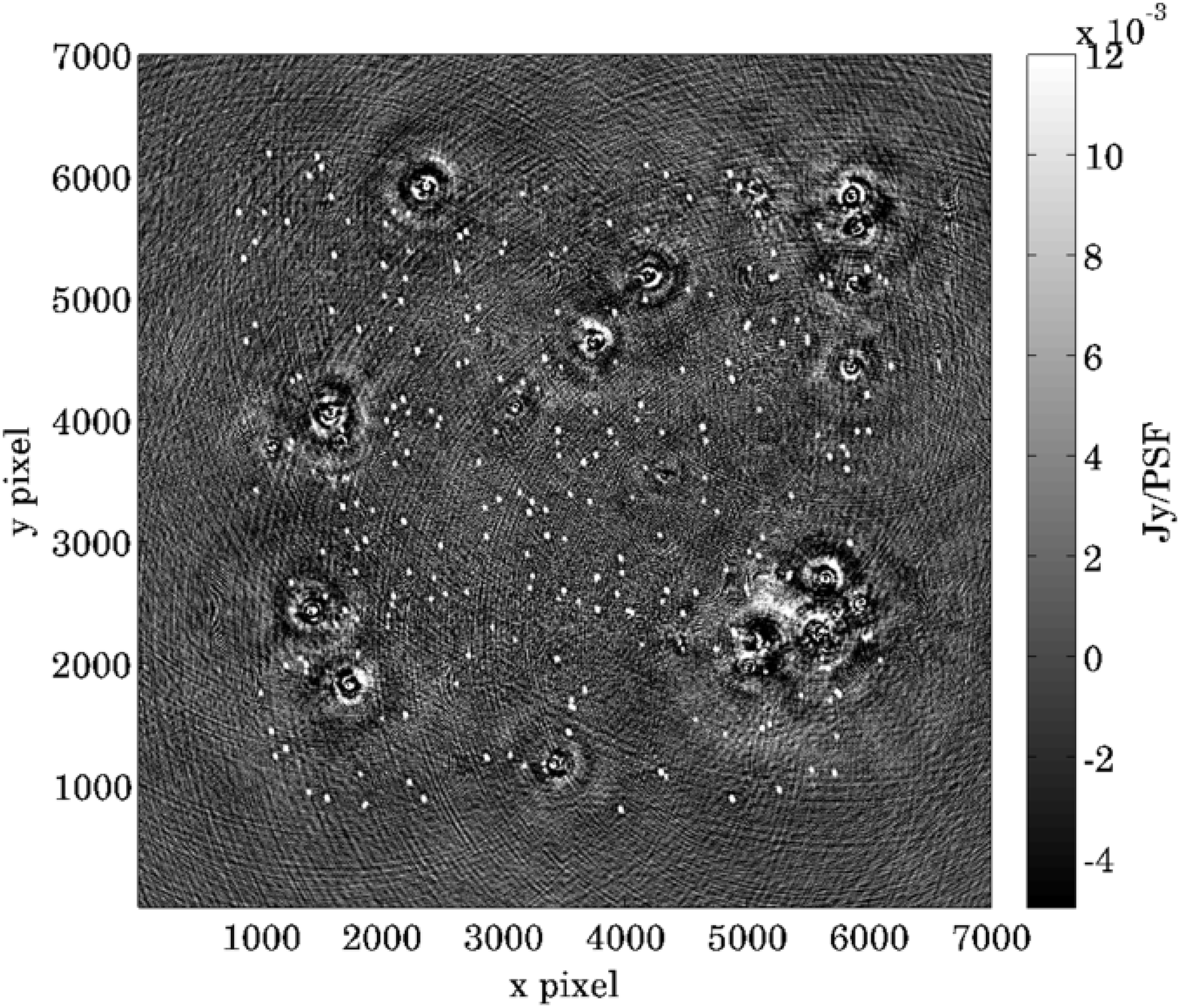,width=8.0cm}}
\vspace{0.5cm} \centerline{(a)}\smallskip
\end{minipage}\\
\begin{minipage}{0.48\linewidth}
\centering
 \centerline{\epsfig{figure=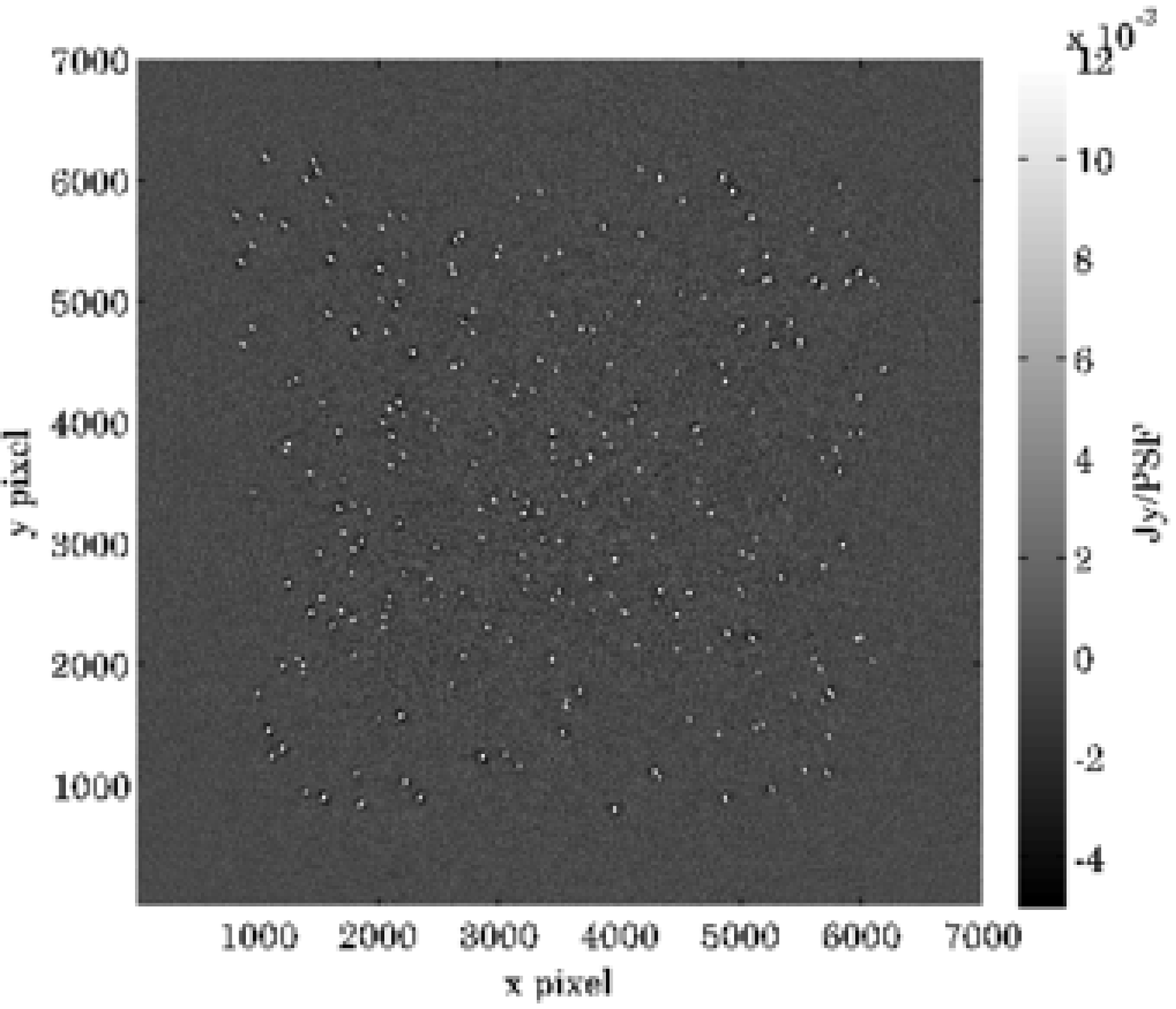,width=8.0cm}}
\vspace{0.5cm} \centerline{(b)}\smallskip
\end{minipage}
\begin{minipage}{0.48\linewidth}
\centering
 \centerline{\epsfig{figure=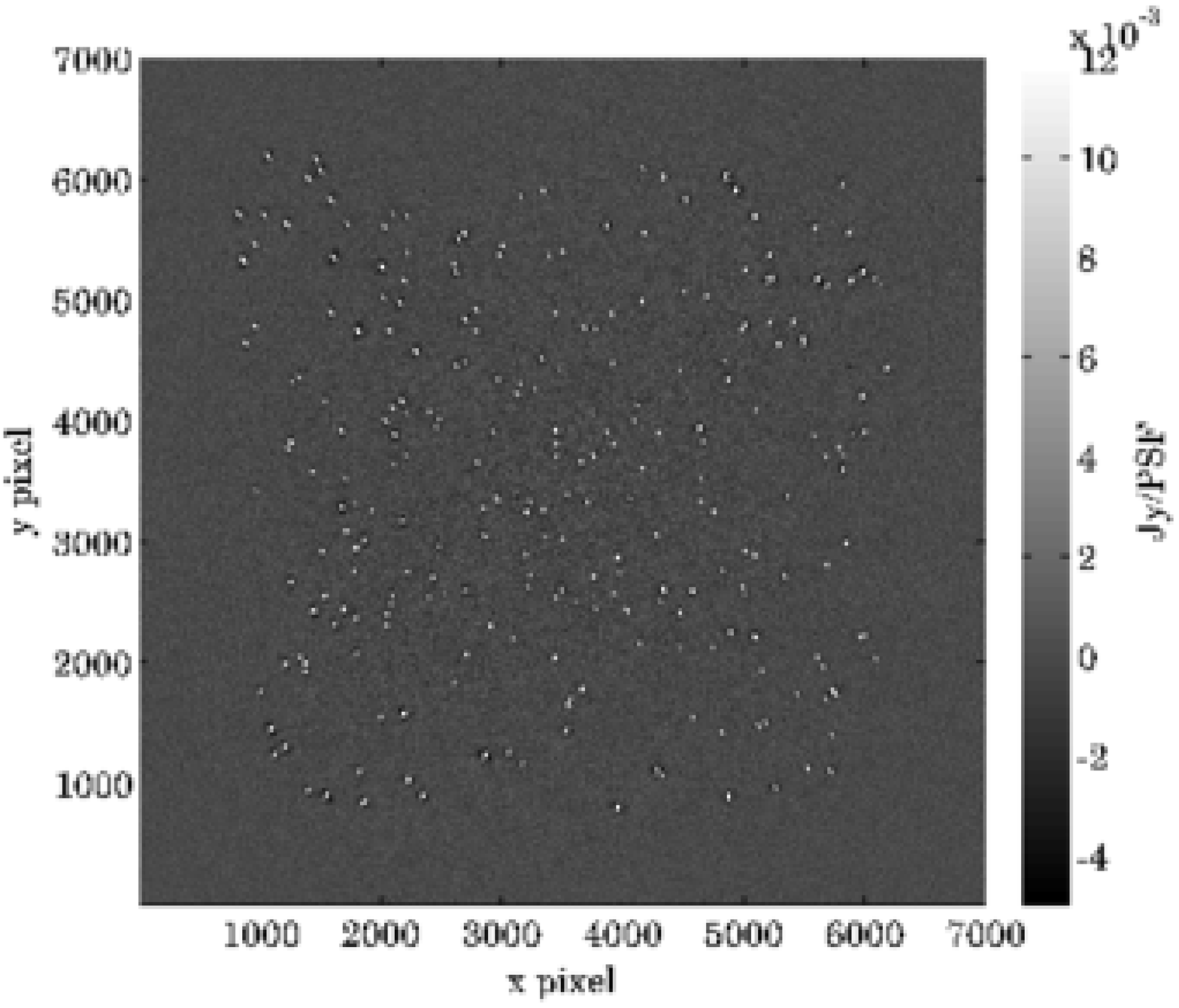,width=8.0cm}}
\vspace{0.5cm} \centerline{(c)}\smallskip
\end{minipage}
\end{center}
\caption{Continuum images of simulated data with $K=25$ point sources with errors across a field of view of about 7$\times$7 square degrees. (a) Raw image before calibration (b) Residual image after normal calibration (c) Residual image after distributed calibration. Both (b) and (c) are visually similar but (c) has lower noise.\label{sim_k25}}
\end{minipage}
\end{figure*}

In this simulation we set $K=25$ and we use 20 time samples in calibration, in other words, calibration is performed for every 200 s of data. Therefore, for a 6 hour observation, calibration is performed 108 times.  We use $F=2$ and $\rho=5$ and each calibration uses 20 C-ADMM iterations. In each C-ADMM iteration, there are 3 SAGE iterations. In Fig. \ref{sim_k25} (a) we show the uncalibrated continuum image which is dominated by the errors along strong sources. In Figs. \ref{sim_k25} (b) and \ref{sim_k25} (c) we show the calibrated image after normal calibration and distributed calibration, respectively. The noise (at the edge) in Figs. \ref{sim_k25} (a), (b), and (c) are 3.3 mJy, 0.64 mJy and 0.49 mJy, respectively. Therefore, there is a clear reduction in noise with distributed calibration, although this is not visible in Fig. \ref{sim_k25}. 

\begin{figure*}
\begin{minipage}{0.98\linewidth}
\begin{center}
\begin{minipage}{0.48\linewidth}
\centering
 \centerline{\epsfig{figure=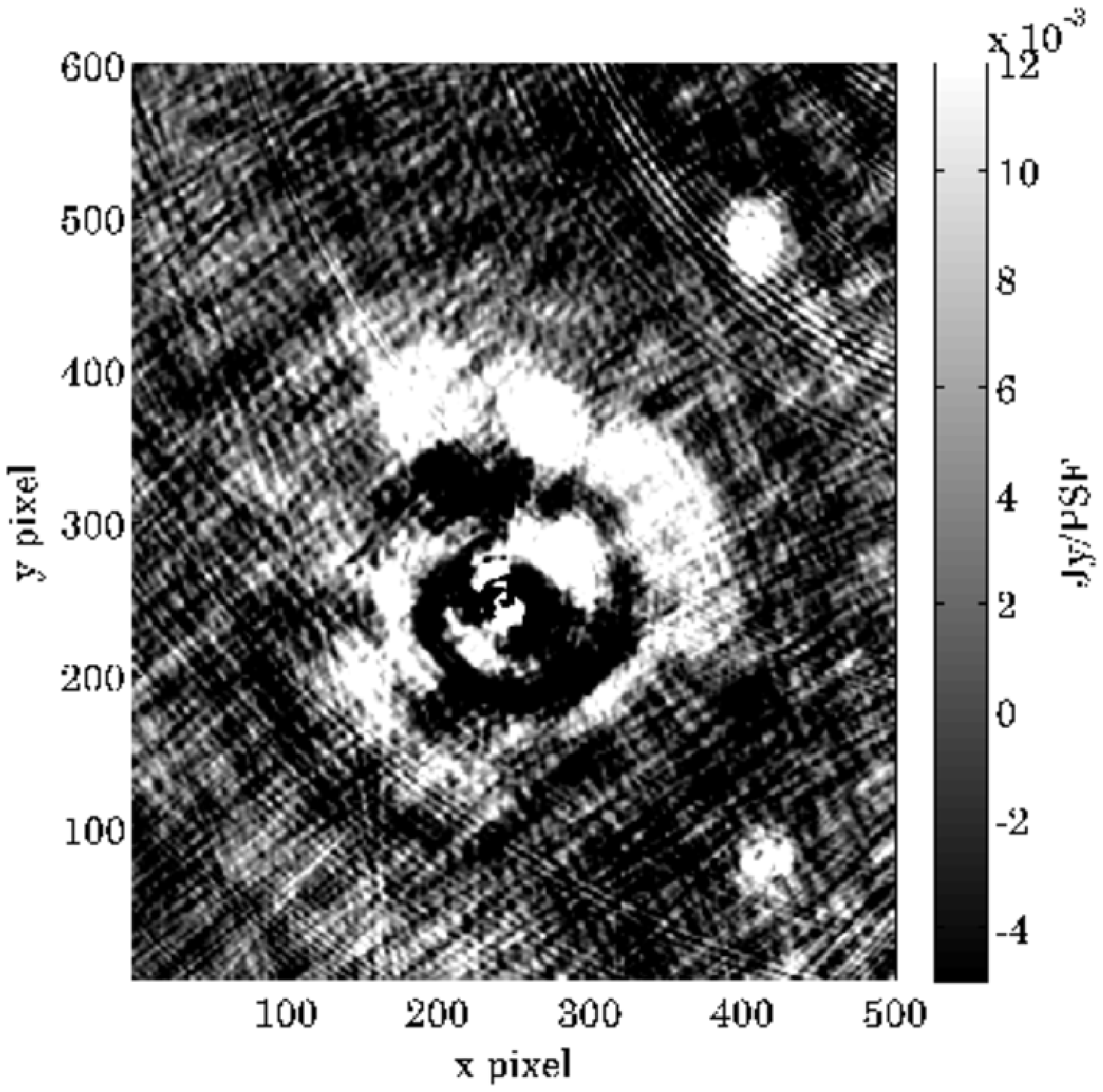,width=8.0cm}}
\vspace{0.5cm} \centerline{(a)}\smallskip
\end{minipage}\\
\begin{minipage}{0.48\linewidth}
\centering
 \centerline{\epsfig{figure=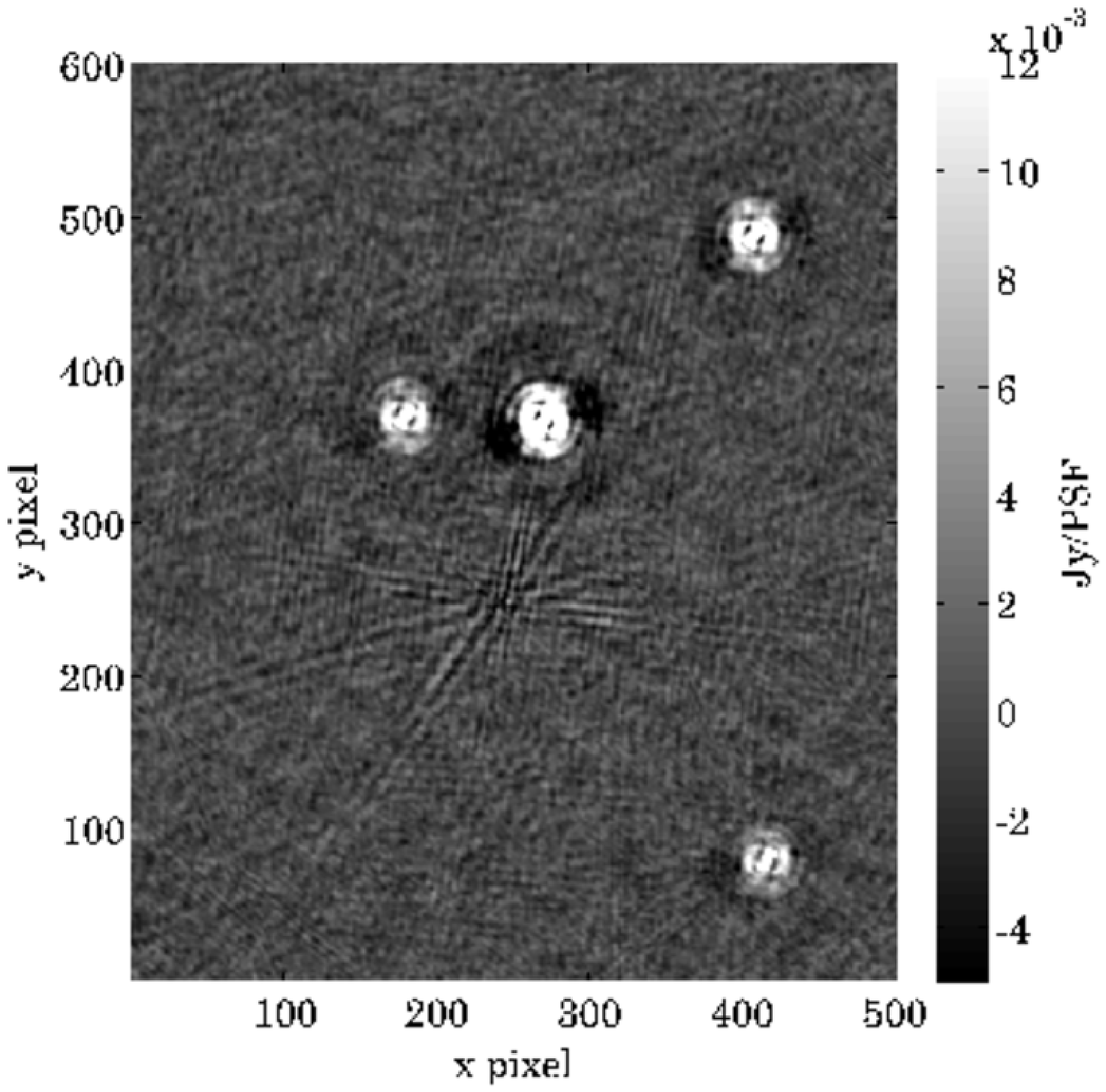,width=8.0cm}}
\vspace{0.5cm} \centerline{(b)}\smallskip
\end{minipage}
\begin{minipage}{0.48\linewidth}
\centering
 \centerline{\epsfig{figure=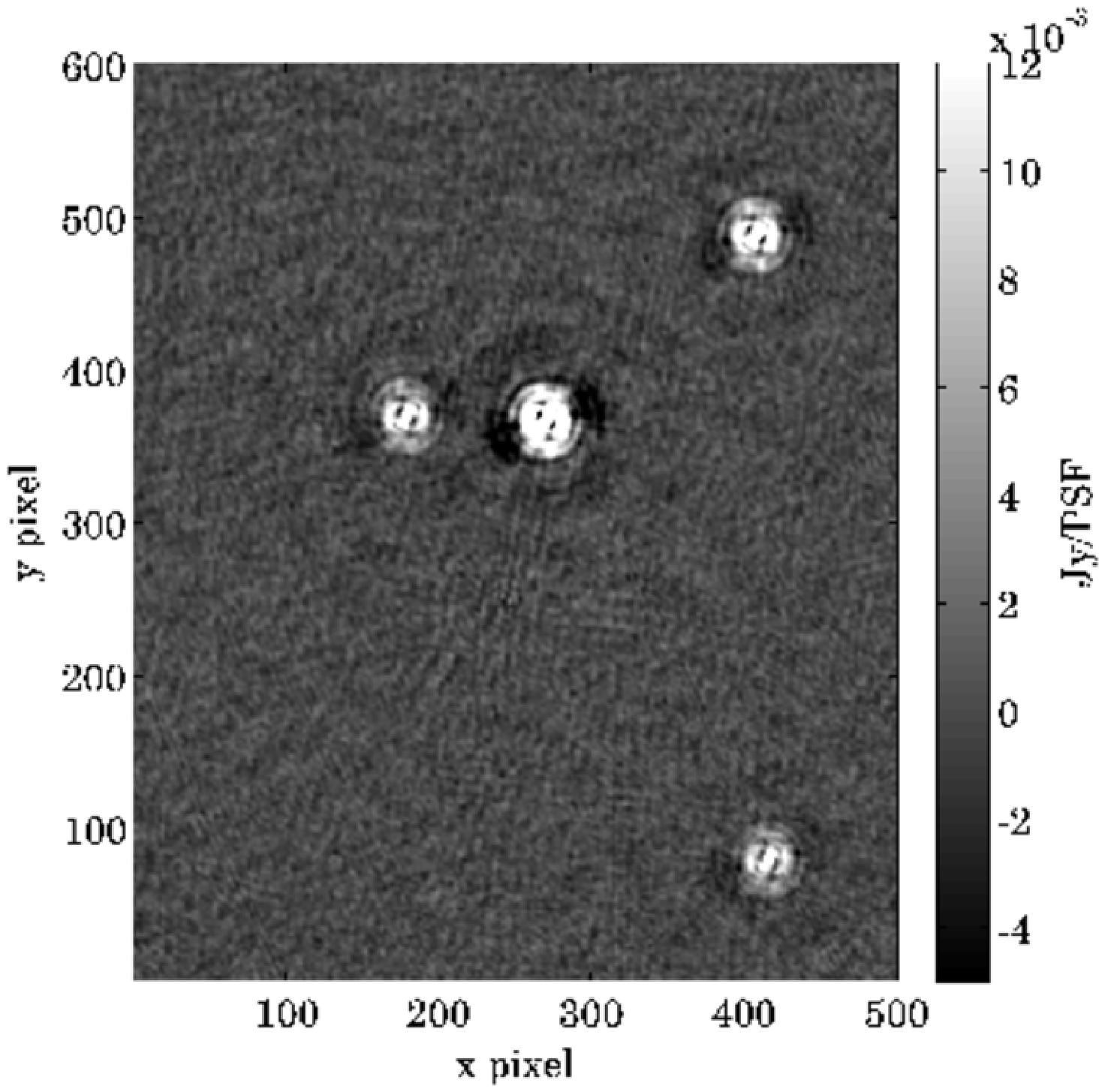,width=8.0cm}}
\vspace{0.5cm} \centerline{(c)}\smallskip
\end{minipage}
\end{center}
\caption{A small area of Fig. \ref{sim_k25} showing a bright source. (a) Raw image before calibration (b) Residual image after normal calibration (c) Residual image after distributed calibration. In both (b) and (c) the errors due to the bright source have mostly disappeared, but (b) has a spike like error pattern that is also removed in (c).\label{sim_k25_zoom}}
\end{minipage}
\end{figure*}

In order to clearly show the difference, we have shown a small area of the full image in Fig. \ref{sim_k25_zoom} where we show an area surrounding a bright source. The uncalibrated image is shown in Fig. \ref{sim_k25_zoom} (a) and images after normal and distributed calibration are shown in Fig. \ref{sim_k25_zoom} (b) and Fig. \ref{sim_k25_zoom} (c), respectively. It is clear that both normal and distributed calibration does well in removing the source, and making the weak sources clearly visible. However, in Fig.  \ref{sim_k25_zoom} (b), there still is an error pattern at the location of the bright source. The magnitude of this error pattern is far below the noise floor of a single channel. Therefore, it is impossible to eliminate this error by normal calibration. However, as seen in Fig. \ref{sim_k25_zoom} (c), distributed calibration does a much better job in removing this error pattern. This also explains the reduction of noise in Fig. \ref{sim_k25} (c). We see similar error patterns in real observations \citep{NCP2013}, and with distributed calibration, the quality of images can certainly be improved.

\section{Conclusions}\label{sec:conclusions}
We have proposed consensus optimization as a way of performing radio interferometric calibration in a distributed way. Distributed calibration enables us to improve the quality of calibration { as well as to distribute the overall computational cost}. The aspect we used for consensus is the smoothness of calibration parameters over frequency. However, a similar strategy can also be adopted to exploit spatial and temporal smoothness that can be explored in future work. We have given simulation results to confirm the feasibility of distributed calibration and also the expected improvement in performance, for instance by avoiding converging to local minima in the optimization. Future work would address better interpolation schemes that enforce consensus { as well as multiplexing schemes when the number of frequency channels that needs to be calibrated is higher than the number of available compute agents}. The source code for the algorithms described in this paper is available at http://sagecal.sf.net/.

\section*{Acknowledgments}
We thank the referee, Yves Wiaux, for the careful review and valuable comments.
This work was financially supported by NWO (grant TOPGO 614.001.005)
and the European Research Council (project LOFARCORE, grant \# 339743).

\bibliographystyle{mn2e}
\bibliography{references}
\bsp
\label{lastpage}
\begin{appendix}
\section{Gradient and Hessian Calculation}\label{app:grad}
This section describes the derivation of the gradient and Hessian operators for (\ref{sum3}) and (\ref{AugL}) so that the Riemannian trust-region algorithm \citep{RTR} can be applied. Without loss of generality, we drop the superscript $f_i$ in this section. We consider ${\bf {\sf J}}$ to be on a matrix manifold denoted by $\mathcal{M}$. Let the function to be minimized be $g({\bf {\sf J}})$. We define the inner product for two elements in the tangent space ${\mathcal T}\mathcal{M}$ of this manifold as
\beq \label{inprod}
h({\bmath \xi},{\bmath \eta})\buildrel\triangle\over= \mathrm{trace}({\bmath \xi}^H{\bmath \eta} +{\bmath \eta}^H{\bmath \xi}),\ {\bmath \xi},{\bmath \eta} \in {\mathcal T}\mathcal{M}.
\eeq
With this definition, the gradient is calculated to satisfy 
\beq \label{grad}
h({\bmath \xi},\mathrm{grad}(g({\bf {\sf J}})))=Dg({\bf {\sf J}})[{\bmath \xi}],\ \forall {\xi} \in {\mathcal T}\mathcal{M}
\eeq
where 
\beq
Dg({\bf {\sf J}})[{\bmath \xi}] \buildrel\triangle\over= \underset{\tau\rightarrow 0}{\mathrm{lim}} \frac{g({\bf {\sf J}}+\tau {\bmath \xi})-g({\bf {\sf J}})}{\tau}.
\eeq
Similarly, the Hessian of $g({\bf {\sf J}})$ can be obtained as
\beq \label{hess}
\mathrm{Hess}\ g({\bf {\sf J}})[{\bmath \eta}]\buildrel\triangle\over =  \underset{\tau \rightarrow 0}{\mathrm{lim}} \frac{1}{\tau}\left( \mathrm{grad}\ g({\bf {\sf J}}+\tau{\bmath \eta})-\mathrm{grad}\ g({\bf {\sf J}})\right).
\eeq
Now we can rewrite (\ref{sum3}) as
\beqn \label{sumtrace1}
\lefteqn{g({\bf {\sf J}})=}\\\nonumber
&&\sum_{p,q} \mathrm{trace}\left(\left({\bf {\sf V}}_{pq} - {\bf {\sf A}}_p {\bf {\sf J}} {\bf {\sf C}}_{pq} \left({\bf {\sf A}}_q {\bf {\sf J}}\right)^H\right)^H \right. \\\nonumber
&& \mbox{} \times \left. \left({\bf {\sf V}}_{pq} - {\bf {\sf A}}_p {\bf {\sf J}} {\bf {\sf C}}_{pq} \left({\bf {\sf A}}_q {\bf {\sf J}}\right)^H\right) \right)
\eeqn
and we can rewrite (\ref{AugL}) as
\beqn \label{sumtrace2}
\lefteqn{L_i\left({\bf {\sf J}},{\bf {\sf Z}},{\bf {\sf Y}}\right)}\\\nonumber
&& = g({\bf {\sf J}}) + \frac{1}{2} \mathrm{trace}\left({\bf {\sf Y}}^H ({\bf {\sf J}}- {\bf {\sf B}} {\bf {\sf Z}}) + ({\bf {\sf J}}- {\bf {\sf B}} {\bf {\sf Z}})^H {\bf {\sf Y}}\right)\\\nonumber
&& + \frac{\rho}{2} \mathrm{trace}\left(\left({\bf {\sf J}}- {\bf {\sf B}}{\bf {\sf Z}}\right)^H \left({\bf {\sf J}}- {\bf {\sf B}}{\bf {\sf Z}}\right)\right).
\eeqn
Finally, applying (\ref{grad}) and (\ref{hess}) to (\ref{sumtrace1}) and (\ref{sumtrace2}) gives us  (\ref{gradL}) and (\ref{HessL}). Moreover, by taking gradient with respect to ${\bf {\sf Z}}$, we also get (\ref{gradZ}). { Note also that since we use the RTR algorithm in Euclidean space, the projection is $\Pi({\bf {\sf J}})={\bf {\sf J}}$ and the retraction is $R({\bf {\sf J}},{\bmath \eta}) = {\bf {\sf J}} + {\bmath \eta}$.
}

\section{Convergence}\label{app:conv}
First, we consider the convergence of the RTR algorithm in minimizing a function such as $g({\bf {\sf J}})$ in (\ref{sum3}) with respect to ${\bf {\sf J}}$. Using  \citep[7.4.6]{AMS}, we only need to show that the manifold on which ${\bf {\sf J}}$ lies (say $\mathcal{M}$) is compact (smoothness of $g({\bf {\sf J}})$ is obvious). Given that the sky model has finite and non-zero flux and the data has finite power, we see that $\|{\bf {\sf J}}\|$ is finite and hence $\mathcal{M}$ is bounded. 

Each pair of $p,q$ in (\ref{sum3}) gives us a set of constraints on the values of ${\bf {\sf J}}$ that can be expressed as a set of nonlinear functions $\widetilde{g}_{pq,ij}(.,,.\ldots)=0$ for different values of $p,q,i,j$, which are actually mappings from $\mathbb{R}^{8N}$ to $\mathbb{R}$. Since $0$ is a closed set, elements of ${\bf {\sf J}}$ are in the inverse image of $\widetilde{g}_{pq,ij}(.,,.\ldots)=0$ which is also a closed set. Note that in order to have expressions such as $\widetilde{g}_{pq,ij}(.,,.\ldots)=0$ that are unique, we need to have at least few of them with nonzero values for ${\bf {\sf C}}_{pq}$ { and unique $uv$ coordinates}. In other words, we need to have a sky model with non zero total flux. For all possible values of $p,q,i,j$, with unique $\widetilde{g}_{pq,ij}(.,,.\ldots)=0$ expressions, we get an intersection of closed sets within which the elements in ${\bf {\sf J}}$ must lie. Since the intersection of closed sets gives a closed set, we see that $\mathcal{M}$ on which ${\bf {\sf J}}$ lies is also closed. Therefore $\mathcal{M}$ is both bounded and closed and by Heine-Borel theorem \citep{AMS}, it is compact.

For the convergence of the C-ADMM algorithm, we need to have smooth, convex functions for $g({\bf {\sf J}})$. This is not always guaranteed but with same assumptions as above, most of the time it can be safely assumed to be convex (see \cite{2012ExA} for a detailed investigation).
\end{appendix}
\end{document}